\documentclass[aip, cha, amsmath, amssymb, reprint]{revtex4-1}
\usepackage{graphicx}
\usepackage{dcolumn}
\usepackage{bm}
\usepackage{lipsum}
\usepackage{multirow}
\usepackage{color}

\definecolor{Myorange}{cmyk}{0,0.42,1,0}

\begin{document}
\preprint{AIP/123-QED}
\title{Detection of time reversibility in time series by ordinal patterns analysis}

\author{J. H. Mart\'inez}
\email{johemart@gmail.com}

\affiliation{INSERM-UM1127, Sorbonne Universit\'e, Institut du Cerveau et de la Moelle Epini\`ere. France}%

\author{J. L. Herrera-Diestra}%
\affiliation{ICTP South American Institute for Fundamental Research, IFT-UNESP. Brazil}

\author{M. Chavez}
\affiliation{CNRS UMR7225, H\^opital Piti\'e Salp\^etri\`ere. France}%

\date{\today}
\begin{abstract}
Time irreversibility is a common signature of nonlinear processes, and a fundamental property of non-equilibrium systems driven by non-conservative forces. 
A time series is said to be reversible if its statistical properties are invariant regardless of the direction of time. Here we propose the Time Reversibility from Ordinal Patterns method (TiROP) to assess time-reversibility from an observed finite time series. TiROP captures the information of scalar observations in time forward, as well as its time-reversed counterpart by means of ordinal patterns. The method compares both underlying information contents by quantifying its (dis)-similarity via Jensen-Shannon divergence. The statistic is contrasted with a population of divergences coming from a set of surrogates to unveil the temporal nature and its involved time scales.
We tested TiROP in different synthetic and real, linear and non linear time series, juxtaposed with results from the classical Ramsey's time reversibility test.
Our results depict a novel, fast-computation, and fully data-driven methodology to assess time-reversibility at different time scales with no further assumptions over data. 
This approach adds new insights about the current non-linear analysis techniques, and also could shed light on determining new physiological biomarkers of high reliability and computational efficiency.
\end{abstract}

\maketitle

\begin{quotation}
Most time series observed from real systems are inherently nonlinear, thus detecting this property is of full interest in natural or social sciences. 
One feature that ensures the nonlinear character of a system is the time irreversibiliity. A time series is said to be reversible if its statistical properties are invariant regardless of the direction of time. Here we propose the Time Reversibility from Ordinal Patterns (TiROP) method to assess the temporal symmetry of linear and nonlinear time series  at different scales. Our approach is based on a fast-computing symbolic representation of the observed data. Here, TiROP is compared with a classical time-reversibility test in a rich variety of synthetic and real time series from different systems, including ecology, epidemiology, economy and neuroscience.
Our results confirm that TiROP has a remarkable performance at unveiling the time scales involved in the temporal irreversibility of a broad range of processes.
\end{quotation}

\section{Introduction}
A time series is said to be reversible if its statistical properties are invariant regardless of the direction of time. Time irreversibility is a fundamental property of non-equilibrium systems~\cite{Lamb1998, Prigogine1999, Andrieux2007} and dynamics resulting from non-conservative forces (memory)~\cite{Puglisi2009}, therefore, it is expected to be present in the scalar observation of different biological and physical systems. Indeed, time irreversibility has been reported in ecological and epidemiological time series~\cite{Grenfell1994, Stone1996}, in tremor time series of patients with Parkinson's disease~\cite{Timmer1993}, in electroencephalographic (EEG) recordings of epileptic patients~\cite{Heyden1996, Pijn1997, Schindler2016}, or in cardiac interbeat interval time series extracted from patients and healthy subjects under different cardiac conditions~\cite{Costa2005, Porta2008, Casali2008, Porta2009}.

Any time series that is a realisation of a stationary, linear Gaussian process is time reversible, because of the symmetry of their covariance functions~\cite{Cox1981, Hallin1988, Lawrance1991}. Nevertheless, a non-Gaussian amplitude distribution could be due to a static nonlinear transformation of a stationary linear Gaussian process, and by itself is no proof of temporal irreversibility. Furthermore, non-Gaussian processes modeled as outputs of linear systems are reversible~\cite{Weiss1975}. In contrast, the output of a non-linear system excited by non-Gaussian noises is time irreversible~\cite{Rao1998}. Non-linear and non-Gaussian linear models typically have temporal directionality as a property of their higher-order dependency~\cite{Lawrance1991}. The study of time reversibility properties of time series might therefore provide meaningful insights into the underlying nonlinear mechanisms of the observed data.

Classical time reversibility tests require higher-order moments of the studied signal $X_t$ to be finite~\cite{Pomeau1982, Ramsey1988, deLima1997}. Other tests have been devised by directly comparing the distribution of vectors~\cite{Diks1995, Heyden1996, Porporato2007} $\{X_{t}, X_{t+1}, \cdots,X_{t+D}\}$ and its time-reversed version $\{X_{t+D}, X_{t+D-1},\cdots,X_{t}\}$, or from the projection of dynamics onto a finite number of planes~\cite{Guzik2006, Casali2008}. In the last years, some works have proposed statistical tests for irreversibility based on the so-called visibility graphs~\cite{Lacasa2012}, i.e., the mutual visibility relationships between points in a one-dimensional landscape representing $X_t$~\cite{Donges2013, Flanagan2016, Li2018}. These works show that irreversible dynamics results in an asymmetry between the probability distributions of graph properties (e.g. links or paths-based characteristics). Recently, this approach has been extended for the study of non-stationary processes~\cite{Lacasa2015, Li2018b}. 

For real-valued time series, some studies have  proposed time-reversibility tests based on different symbolization procedures to characterize the dimensional phase spaces of $X_t$ and its time-reversed version~\cite{Daw2000, Costa2005, Zanin2018}. These symbolic transformations are generally done by defining a quantization procedure to transform the time series into a discrete sequence of unique patterns or symbols~\cite{Kennel2004, Porta2008, Porta2009}. Some of these reversibility tests use a priori binomial statistics to assess statistical significance of findings~\cite{Daw2000, Zanin2018}. Nevertheless, such tests assume independence of the observed symbols, which is unlikely to occur in real data with temporal correlations. In case of such serial correlations, a rigorous theoretical framework cannot be derived and Monte Carlo simulations (e.g. parametric or non-parametric re-sampling) must be performed to estimate the significance level of time reversibility tests~\cite{Costa2005, Porta2008, Porta2009}.

In this work we propose a novel procedure, the Time Reversibility from Ordinal Patterns method (TiROP), that compares the empirical distributions of the forward and backward statistics of a time series. To estimate the asymmetry between both probability distributions we use the ordinal symbolic representations~\cite{Band2002, Amigo2015}. In contrast with other approaches based on symbolic analysis, the ordinal patterns analysis used here is fully data-driven, i.e., the symbolic transformation does not require any \textit{a priori} threshold, or any knowledge about the data sequence. We complete our time reversibility test with surrogate data analysis without making assumptions on the underlying generating process~\cite{Schreiber1996, Schreiber2000}.

The proposed framework is validated on synthetic data simulated with linear, nonlinear, non-Gaussian stochastic and deterministic processes. The method is also illustrated on a collection of different real time series. The reliability and performances of our method are also compared with those obtained by a classical moment-based method. The remainder of the paper is organized as follows: Section II describes the proposed framework, as well as the comparative method used to benchmark our solution. Experimental results and evaluation of the method in synthetic time series are in Section III; while the evaluation of the test on real data is provided in Section IV. Finally, we conclude the paper with a discussion in Section V.

\begin{figure}
\includegraphics[width=0.5\textwidth]{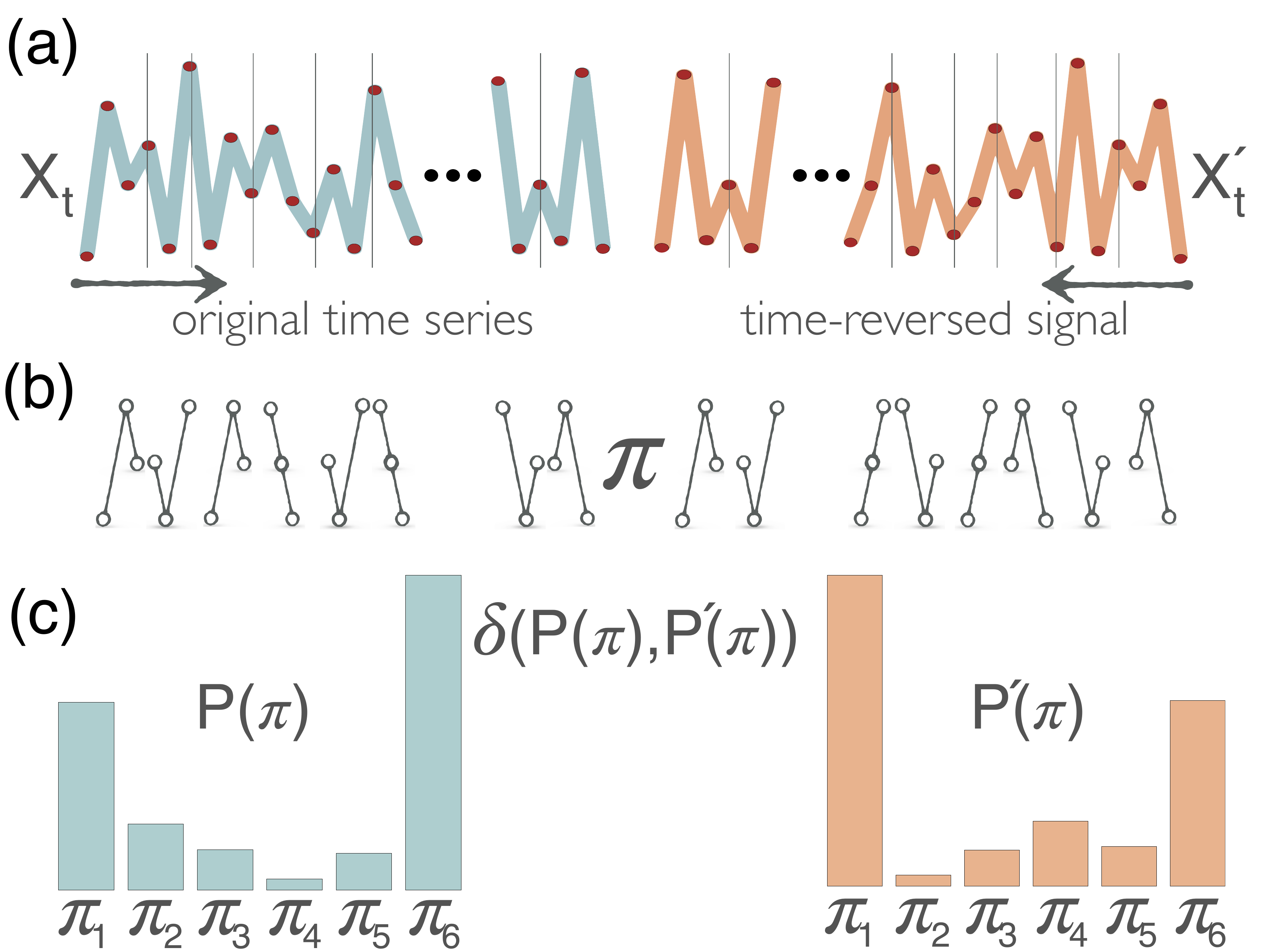}
\caption{\label{fig01} Main steps of the TiROP algorithm for evaluating the time-reversibility of a time series $X_t$. (a) (Left) Original time series represented in blue. (Right) The time reversed signal $X'_t$ represented in orange. (b) Patterns $\pi$'s extracted from $X_t$ and $X'_t$ for $D=3$. (c) Probability distributions $P(\pi)$ and $P'(\pi)$ extracted from $X_t$ and $X'_t$, respectively. The Jensen-Shannon $\delta$ captures the dissimilarity between the information content in both distributions}
\end{figure}

\section{Methods}
\subsection*{Capturing information dynamics from time series}
Symbolisation procedures map a time series $X_t$ onto a discretized symbols sequence by extracting its amplitudes' information~\cite{AmigoBook}. Among several symbolisation proposals \cite{Daw2003}, we considered here the dynamical transformation by Band and Pompe~\cite{Band2002}. This method maps a time series $X_t$ with $t = 1,\ldots,T$ to a finite number of patterns that encode the relative amplitudes observed in the $D$-dimensional vector $\mathbf{X}_t=\{X_t, X_{t+\tau}, \ldots, X_{t+(D-1)\tau}\}$. The elements of the vector $\mathbf{X}_t$ are mapped uniquely onto the permutation $\mathbf{\pi}  =(\pi_0,\pi_1,\ldots,\pi_{D-1})$ of $(0,1,\ldots,D-1)$ that fulfills $X_{t+\pi_0\tau} \leqslant X_{t+\pi_1\tau}, \leqslant \ldots \leqslant X_{t+\pi_{D-1}\tau}$. Each order pattern (permutation) represents thus a subset of the whole embedding state space.

The set of all possible ordinal patterns derived from a time series is noted as $S_t$, whose cardinality is $D!$ at most. The whole sequence of ordinal patterns extracted from $X_t$ is known as the symbolic representation of the time series. The information content of $X_t$ is captured by the probability density $P(\pi)$ of finding a particular pattern of order $D$ in $S_t$. The higher the order is, the more information is captured from the time series. To sample the empirical distribution of ordinal patterns densely enough for a reliable estimation of its probability distribution we follow the condition~\cite{Amigo2007} $T\geqslant (D+1)!$

\begin{table*}
\caption{\label{tab01} Synthetic models. LGP and AR(2) are two linear reversible processes. The non-linear (non-reversible) AR models are driven by a Laplacian and bimodal noise distribution, respectively. Two Self-Exciting Threshold AutoRegressive models, SETAR(2; 2,2) and SETAR(2; 3,2), are non linear models with regime switching behavior. The last two models, R\"ossler and Lorenz oscillators, are set under chaotic regime.}
\begin{ruledtabular}
\begin{tabular}{lc} 
 Model & Equation \\ \hline \vspace{0.1cm}
LGP & Gaussian noise with distribution $\mathcal{N}(0,1)$ \\ \vspace{0.2cm}
AR(2)\footnote{$\epsilon_t$ denotes white noise processes.}    & $x_{t+2} = 0.7x_{t+1} + 0.2x_t + \epsilon_t$ \\ \vspace{0.1cm}
 \multirow{2}{*}{N-AR(2)\footnote{noises $\{\eta'_t$,$\eta''_t\}$ are iid. See main text for the parameters}} & $x_t = 0.5x_{t-1} - 0.3x_{t-2} + 0.1y_{t-2} + 0.1x_{t-2}^2 + 0.4y_{t-1}^2 +  0.0025\eta'_t$\\ \vspace{0.2cm}
                                      &  $y_t=\sin(4\pi t)+\sin(6\pi t) + 0.0025\eta''_t$ \\ \vspace{0.2cm}
SETAR(2; 2,2)\footnotemark[1] & $x_t = 
\begin{cases}
0.62 + 1.25x_{t-1}-0.43x_{t-2}+0.0381\epsilon_t & \quad \text{if } x_{t-2} \leq 3.25 \\
2.25 + 1.52x_{t-1}-1.24x_{t-2}+0.0626\epsilon_t & \qquad \text{otherwise}   \\
\end{cases} $ \\ \vspace{0.1cm}
SETAR(2; 3,2)\footnotemark[1]   & $x_t = 
\begin{cases}
0.733 + 1.047x_{t-1}-0.007x_{t-2}+0.242x_{t-3}+0.0357\epsilon_t & \quad \text{if } x_{t-2} \leq 3.083 \\
1.983 + 1.52x_{t-1}-1.162x_{t-2}+0.0586\epsilon_t & \qquad \text{otherwise} \\
\end{cases}$ \\
\multirow{3}{*}{R\"ossler}  & $\dot{x}=-y-z$ \\
											 & $\dot{y}=x+0.2y$ \\ \vspace{0.2cm}
											 & $\dot{z}=0.2+z(x-5.7)$ \\ 
\multirow{3}{*}{Lorenz}  	 & $\dot{x}=10(y-x)$ \\
											 & $\dot{x}=x(28-z)-y$ \\
											 & $\dot{x}=xy-2.6667z$ \\
\end{tabular}
\end{ruledtabular}
\end{table*}

The analysis of ordinal representations has some practical advantages~\cite{Amigo2015}: \textit{i)} it is computationally efficient, \textit{ii)} it is fully data-driven with no further assumptions about the data range to find appropriate partitions and, \textit{iii)} a small $D$ is generally useful in descriptive data analysis~\cite{Band2002}. Furthermore, this symbolisation method is known to be relatively robust against noise, and useful for time series with weak stationarity~\cite{Keller2014, AmigoBook, Cazelles2004, Martinez2018, Zanin2012, Rosso2007}. 

\subsection*{Assessing time reversibility}
A time series $X_t$ is said to be time-reversible if the joint distributions of vectors $\mathbf{X}_t = \{X_{t}, X_{t+1}, \cdots,X_{t+D}\}$ and $\mathbf{X}'_t =\{X_{t+D}, X_{t+D-1},\cdots,X_{t}\}$ for $D$ are equal for all $t$,  i.e., the statistical properties of the process are the same forward and backward in time. All Gaussian processes (and all static transformations of a linear Gaussian process) are time-reversible since their joint distributions are determined by the covariance function which is symmetric~\cite{Weiss1975, Hallin1988}. On the contrary, linear processes driven by non-Gaussian innovations and the nonlinear processes with regime-switching structures, such as the self-exciting threshold autoregressive (SETAR) process~\cite{Tong1980}, are generally time irreversible~\cite{Tong1980, Tong1983, Tong1990, Tong2011}.

Time reversibility implies that the differences of the series being tested have symmetric marginal distributions, i.e. if $X_t$ is time reversible, the distribution of $Y_{t,\tau} = X_{t}- X_{t-\tau}$ is symmetric about the origin for every $\tau$~\cite{Cox1981}. Time reversibility also implies that all the odd moments of $Y_{t,\tau}$, if exist, are zero~\cite{Cox1981}. A simple measure for a deviation from reversibility for a certain time lag $\tau$ was introduced by Ramsey~\cite{Ramsey1988}. Time reversibility is assessed by checking the difference between the sample bi-covariances for zero mean time series $\gamma(\tau) = \langle X_{t}^2 X_{t-\tau}\rangle -  \langle X_{t} X_{t+\tau}^2\rangle$. This method is a benchmark test for time-reversibility and it has been proved to be effective at detecting non-linearity and reversibility in different time series, such as hearth rates, economical data, or even in SETAR models~\cite{Braun1998, Rothman1992, Belaire2003}. Nevertheless, moment-based tests for time reversibility are not really applicable because they require higher-order moments of $X_t$ to be finite, which may rule out many real time series~\cite{deLima1997}. Furthermore, it is quite possible to encounter a situation in which the individual test statistics are significant for some lags but insignificant for others.

\begin{figure*}
\includegraphics[width=1\textwidth]{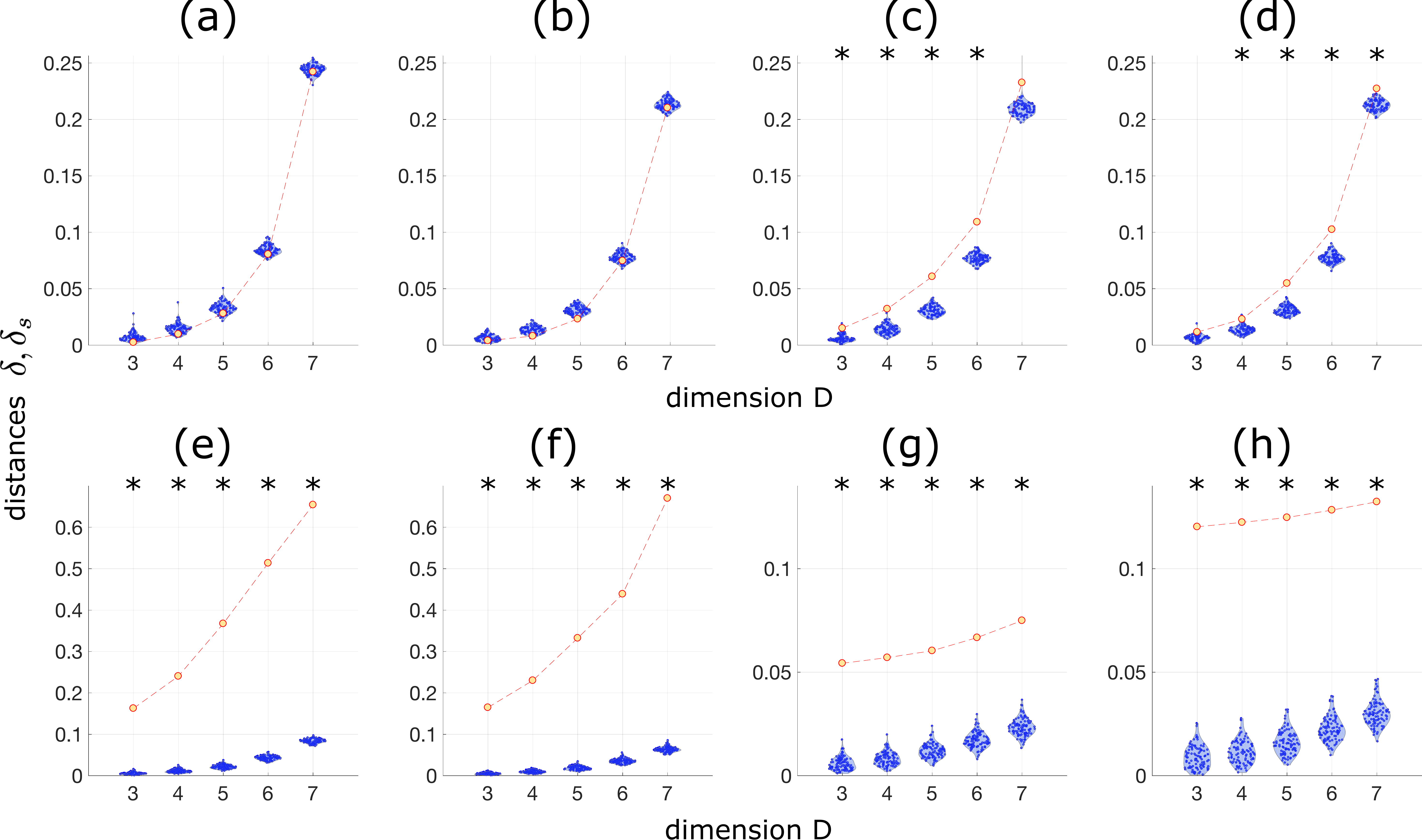}
\caption{\label{fig02} TiROP test applied to synthetic models. Yellow dots indicate the original $\delta$ values for each $D$. Dashed red lines are visual guides but do not represent continuity. Blue dots represent the distributions of $\{\delta_s\}$ at different scales. Black asterisks indicate the dimension $D$ for which the value of $\delta$ is statistically different from $\{ \delta_s \}$. The different models are: (a) Linear Gaussian process; (b) linear autoregressive model; (c) non-linear AR model driven by  a Laplacian noise; (d) non-linear system excited by a noise with a bi-modal distribution; (e) SETAR with  two regimes, each one with second order delays; (f) SETAR model with  two regimes, with delays of third and second order; (g) chaotic R\"ossler system; and (h) chaotic  Lorenz model.}
\end{figure*}

In this work, we propose the Time Reversibility from Ordinal Patterns method (TiROP) as a procedure to assess for time-irreversibility with no assumptions about the process or the observed signal $X_t$ (see the general scheme in Fig.~\ref{fig01}). Ordinal symbolic representations are not symbols ad hoc, but they encode information about the temporal structure of the underlying data. Instead of comparing empirical distributions from $X_t$ and its time-reversed version $X'_t$, we compare the permutation partition (i.e., the symbolic representation) of the embedding state spaces spanned by $\mathbf{X}_t$ and $\mathbf{X}'_t$. The idea behind TiROP is to compare the  distribution $P(\pi)$ of ordinal patterns obtained from the original signal, i.e., the distribution of the ordinal transformation of vectors $\mathbf{X}_t$; with the probability $P'(\pi)$ resulting from its time reversed version $\mathbf{X}'_t$.   

To quantify the (dis)-similarity between both information contents, we use the Jensen-Shannon divergence~\cite{Lin1991} $\delta(P(\pi),P'(\pi))= \frac{1}{2}D(P(\pi),M(\pi))+\frac{1}{2}D(P'(\pi), M(\pi)))$, where $M(\pi) = \frac{1}{2}(P(\pi)+P'(\pi))$ and $D(U,W)=\sum_i U(i)\log \frac{U(i)}{W(i)}$ is the divergence from distribution $U$ to $W$. Time reversibility implies that distributions of vectors $\mathbf{X}_t$ and $\mathbf{X}'_t$, and therefore the distributions of their ordinal transformations, are the same. 

To rule out the possibility that large values of $\delta$ could account for non-Gaussian distributions, or large autocorrelation values at different time lags in signal $X_t$, the statistical significance of $\delta$ values is assessed by a z-test to quantify the statistical deviation from values obtained in an ensemble of surrogate data~\cite{Schreiber2000, Kugiumtzis2002, Barnett2005, Engbert2000}. An ensemble $\{X^s_t\}$ of surrogate time series are created directly from the original dataset through replication of the linear autocorrelation and amplitudes distribution.  In this work, we use the so-called Iterative Amplitude Adjusted Fourier Transform (IAAFT)~\cite{Schreiber1996,Leontitsis2003} that preserves power spectrum density and amplitude distribution of original data, while all other higher-order statistics are destroyed.  For each $X^s_t$, we repeat the procedure of Fig. \ref{fig01} to compute a set of $\{\delta_s\}$ dissimilarities. If the original dissimilarity is statistically distant from the distribution of $\{\delta_s\}$ we can assume that $X_t$ comes from a nonlinear system with a time irreversible dynamics. 

The reliability and performances of our TiROP method are also compared with those obtained by the Ramsey's time-reversibility test $\gamma(\tau)$ based on moments and discussed above~\cite{Ramsey1988}. All significance tests are set at $p<0.05$, Bonferroni-corrected for multiple comparisons (dimensions $D$ or time lags $\tau$). 

\section{Time reversibility in synthetic time series}
In this section, we evaluate the performance of the TiROP method on synthetic time series, simulated with different classes of models (see Table. \ref{tab01}):
\begin{figure*}[!htbp]
\includegraphics[width=1\textwidth]{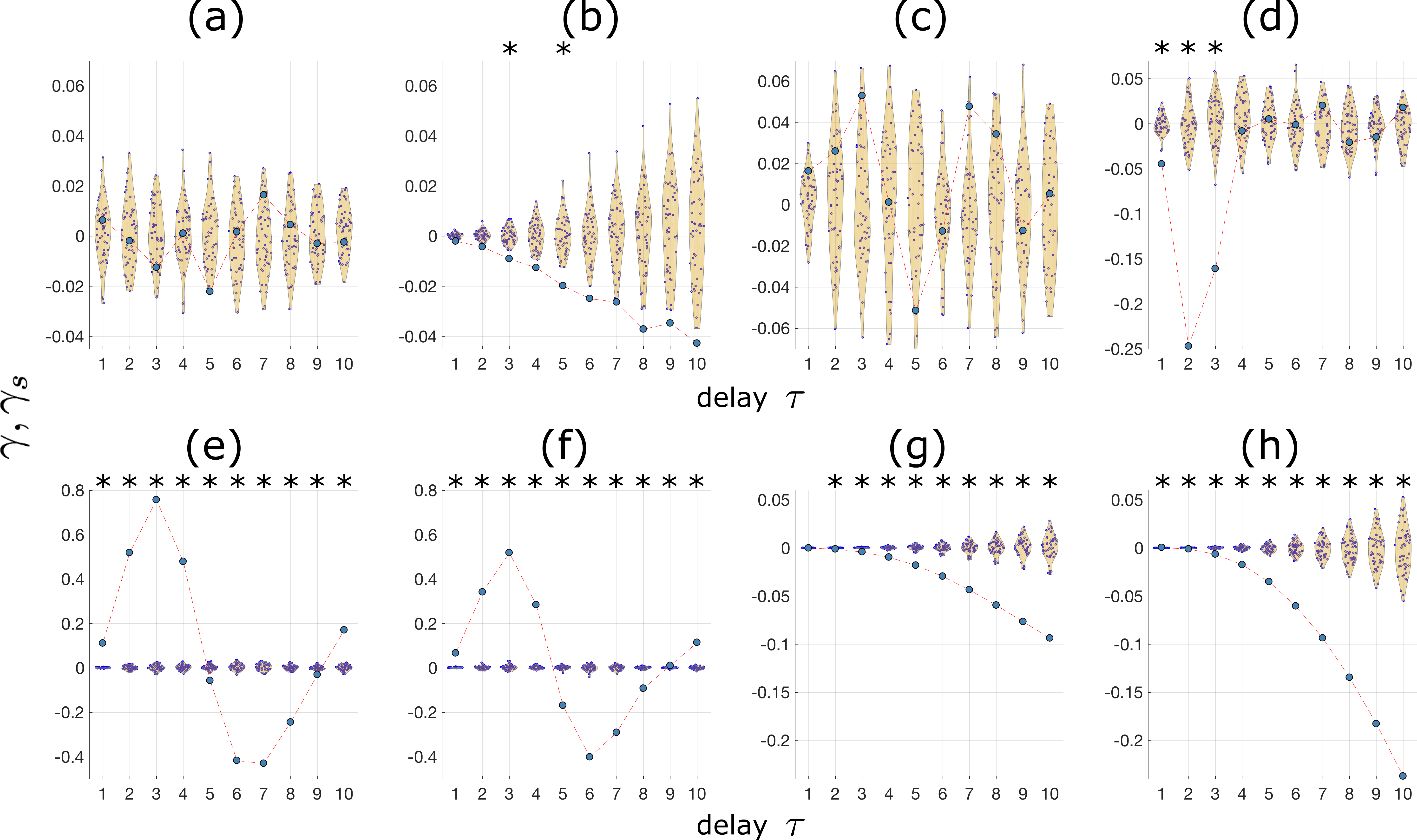}
\caption{\label{fig03} Ramsey's reversibility test of synthetic models. Blue dots indicate the original $\gamma$ values for each time-lag $\tau$. Dashed red lines are visual guides but do not represent continuity. Distributions of $\{\gamma_s\}$ at different scales are represented by the points inside the yellow plots. Black asterisks indicate at which time-lag, $\gamma$ is statistically different from $\gamma_s$. Same stipulations as in the caption of Fig.~\ref{fig02}.}
\end{figure*}

\textbf{Time-reversible linear systems:}  a linear Gaussian process (LGP), and a linear auto-regressive model of second order driven by a white noise.

\textbf{Non-reversible coupled non-linear systems:} Two non-reversible nonlinear AR models (N-AR) driven by non-Gaussian noises~\cite{Rao1998}. We first consider a non-linear system driven by Laplacian noises drawn from the distribution $p(\eta) = \frac{1}{4b}\exp \left(\frac{-|\eta-\mu|}{b}\right)$, with $\mu=0$ and $b=1$. Then, we use the same non-linear model excited by a noise that follows the bi-modal distribution~\cite{Hernandez2016} $p(\eta)=0.5\mathcal{N}(\eta|\mu,\sigma)+0.5\mathcal{N}(\eta|-\mu,\sigma)$, with $\mu=0.63$, $\sigma=1$.

\textbf{Non-reversible switched nonlinear systems:} Two processes from the family of Self-Exciting Threshold AR (SETAR) models, which are largely used to model ecological systems and are characterized for having jumps between different non-linear regimes, each one with different delays~\cite{Tong1983, Petruccelli1990, Rothman1992, Tong2011}. The fifth model is a SETAR with two regimes, each one with second order delays. The sixth one is a SETAR with two regimes with delays of third and second order. 

\textbf{Chaotic, non-reversible systems:} The last two models are the classical R\"osler and  Lorenz systems in their corresponding chaotic regimes . The analyzed time series correspond to the evolution of the $y$ and $z$ variables, from the R\"ossler and Lorenz systems, respectively.

For each model, the length of each time series is set to $T=10^4$, after discarding the first 1000 points to avoid possible transients. Contrary to phase state reconstruction, which requires to select a dimension $D$ and time delay $\tau$ embedding according to some criteria, in ordinal time-series analysis the criteria are computational cost and statistical significance in view of the amount of data available~\cite{Amigo2015, AmigoBook}. We therefore do not make any assumption regarding the dimension, and use different values of $D$ depending on the data length. Although, larger delays can provide additional scale-dependent information about the time series under study, we set $\tau=1$ throughout this work~\cite{Amigo2015, AmigoBook}.  

\begin{figure*}
\includegraphics[width=1\textwidth]{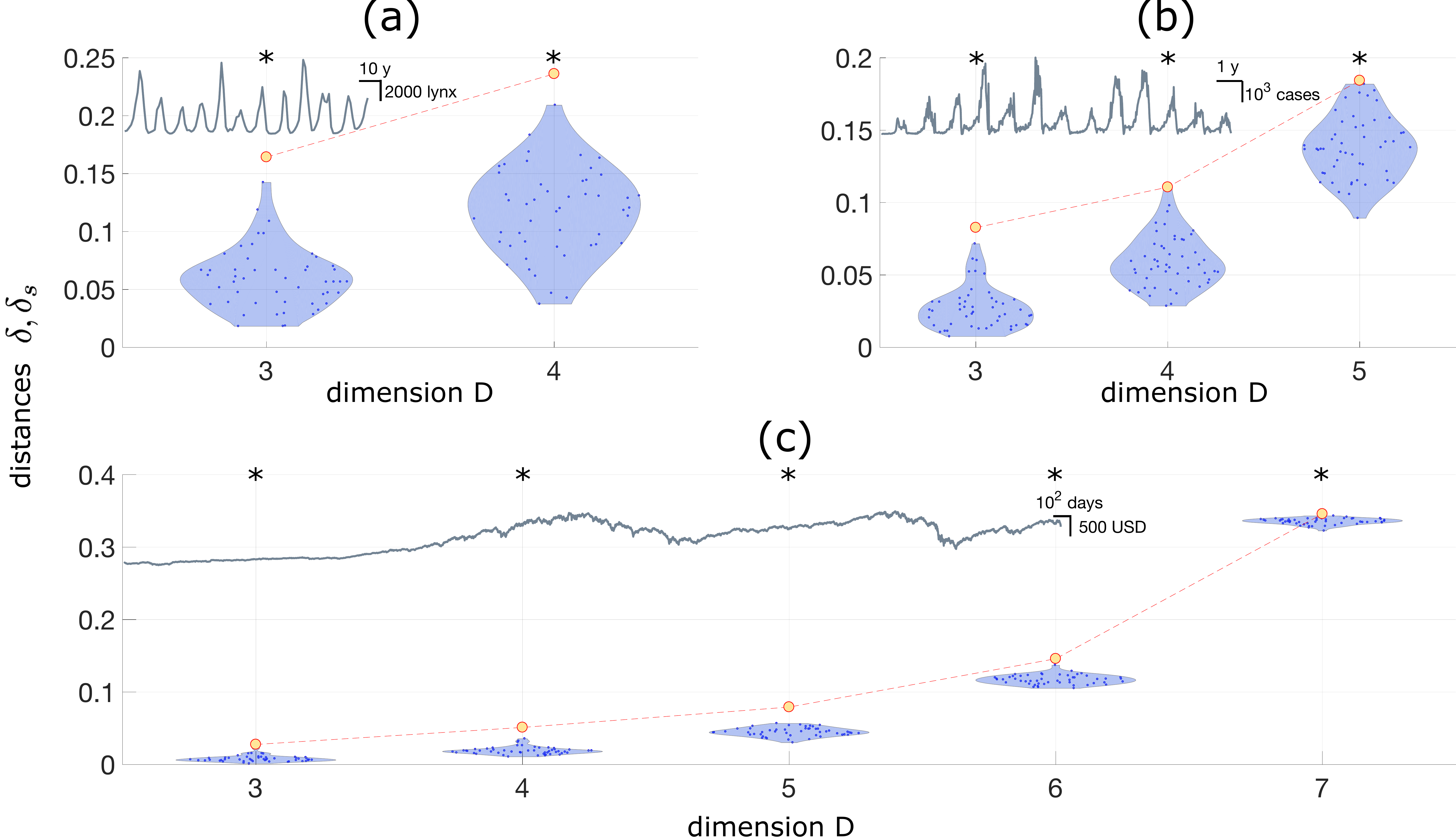}
\caption{\label{fig04} Time-reversibility test on different real data. Insets shows the collection of samples of each process, its temporal and amplitude scales. (a) Time series of lynx returns. (b) Weekly Mexican reported cases of dengue. (c) Daily S\&P closing prices. Yellow dots indicate the original $\delta$ values for each $D$. Dashed red lines are visual guides but do not represent continuity.  Blue dots represent the distributions of $\{\delta_s\}$ at different scales. Black asterisks indicate the dimension $D$ for which the value of $\delta$ is statistically different from $\{ \delta_s \}$.}
\end{figure*}

For the assessment of statistical significance we generate 50 surrogates from each original sequence. For different scales ($D=3,...,7$), we obtain $\delta$ and the set of $\{\delta_s\}$ for all surrogate time series considered. We calculate a $z$-statistics for each $D$ as $|\frac{\delta-\langle \{\delta_s\} \rangle}{\sigma(\{\delta_s\})}|$ and we check for irreversibility by testing the null hypothesis $H_0$ of a time-reversible process with significance level $\alpha\leqslant 0.5$ (corrected by Bonferroni). We repeat the procedure with Ramsey's test, taking into account the first ten time-lags $\tau$'s. Fig. \ref{fig02} (\ref{fig03}) shows the results for TiROP (Ramsey) methodology along different scales (delays).

As expected, the statistical properties of the LPG process are the same forward and backward in time and thus the null hypothesis of reversibility is never rejected by both tests. Interestingly, whereas the TiROP method correctly diagnoses the AR model as a reversible process, Ramsey's statistics yields false positives and falsely rejects $H_0$ in two non-continuous delays.

Whereas non-Gaussian processes modeled as outputs of linear systems are reversible~\cite{Weiss1975}, the output of a non-linear system excited by non-Gaussian noises is time irreversible. For the case of non-linear AR (N-AR) process excited by a Laplacian noise, the null hypothesis of time-reversibility is correctly rejected by our TiROP method, while Ramsey's test fails to detect time-irreversibility along all time-lags. Similar to the previous, the output of the N-AR model driven by a bi-modal noise is detected as irreversible by TiROP for all dimensions $D>1$, while Ramsey's test only detects irreversibility in the first three delays. 

For the SETAR and chaotic models, both TiROP and Ramsey's tests correctly reject the time-reversible hypothesis, in agreement with previous studies at identifying the intrinsic time irreversibility of such models~\cite{Tong1983, Petruccelli1990, Rothman1992, Racine2007, Tong2011}. To notice, however, that Ramsey's statistics yields a false negative at the first delay for the R\"ossler system.

To further evaluate the performance of the TiROP method, we consider short sample sizes. Numerical simulations show that our TiROP test can correctly detected irreversibility in SETAR and chaotic models when the data length is, at least, ten times the fundamental period $T_0$ of SETAR ($T_0 \simeq$ 9 samples) and twelve times the period of chaotic systems ($T_0 \simeq$ 52 samples). For these sample sizes, the Ramsey's method increases dramatically the number of incorrect rejections of true null hypothesis,  as well as the number of false negatives in chaotic systems.

\section{Time reversibility in real data} 
To further demonstrate the potentials of our  test, we apply it to real data of different nature: ecology (the time series of lynx abundance), epidemiology (dengue prevalence), economy (the S\&P price-index series) and neuroscience (electroencephalographic data from an epileptic patient). As data have different length we apply TiROP in different dimensions, following the condition~\cite{Amigo2007} $T\geqslant (D+1)!$

Inset in Fig. \ref{fig04}-(a) shows the well-known time series $x_t$ of fur returns of the Canadian lynx, a valuable collection representing the regularity and rhythm of lynx population in Canada. Each amplitude represents the amount of lynx furs that trappers caught and brought into posts in the same hunting season. $T=114$ samples were collected during 1821-1914 near Mackenzie river region~\cite{Charles1942}. Notice that this dataset was used to fit the SETAR models' parameters used in this work~\cite{Tong1983}. Before applying the time-reversibility test, we applied the variance stabilizing transformation~\cite{Tong1990}  $y_t=\log_{10}(x_t+1)$.  Despite its short data length, our results suggest irreversibility in this time series, in full agreement with previous works~\cite{Petruccelli1990, Lawrance1991, Rothman1992, Tong2011}.

Inset in Fig.~\ref{fig04}-(b) depicts $M=678$ epidemiological weeks of reported cases of Dengue in Mexico during the years 2000-2015~\cite{DengueMXdata}.  As for the lynx time series, time reversibility was assessed on the transformed data $y_t=\log_{10}(x_t+1)$. Based on nonlinear prediction techniques, different studies have proposed evidence for time reversibility in different ecological and epidemiological time series~\cite{Grenfell1994, Stone1996}. For the time series of dengue prevalence considered here, the TiROP method rejects the null hypothesis of time reversibility for all scales. This result indicates that such dengue's dynamics cannot be analyzed by conventional linear models.

The inset in Figure \ref{fig04}-(c) shows $M=5444$ samples from the Standard \& Poor's Index  encompassing the daily historical closing prices from January 1990 to August 2011~\cite{SyP1990}. This is the most representative index of the real situation of market in USA based on the capitalization of 500 large companies with common stocks in NYSE and NASDAQ. Although S\&P-500 time series has been suggested to be irreversible and chaotic~\cite{Vamvakaris2018}, some works have showed that moment-based methods fail at detecting irreversibility~\cite{Racine2007}. To account for the non-stationarity of original data, we extracted the log-returns $y_t=\log(x_{t+1})-\log(x_{t})$, and then we checked for time-reversibility at different scales up to $D=7$. Our method rejects the hypothesis of a time-reversible process, which agrees with previous findings suggesting that irreversibility in economical time series is a rule instead of a simple exception~\cite{Tong1990, Flanagan2016}.

\begin{figure}[!htp]
\includegraphics[width=0.5\textwidth]{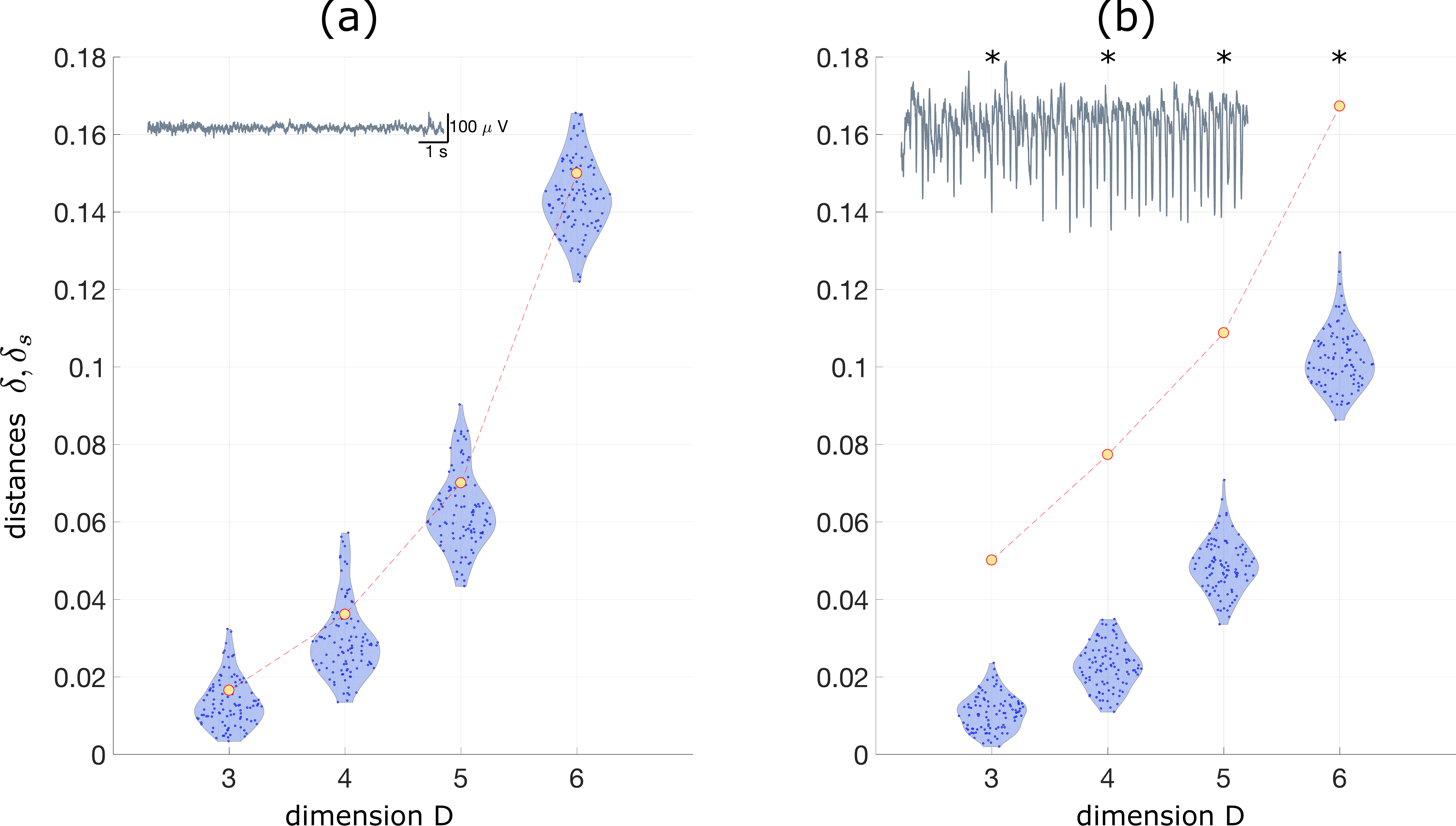}
\caption{\label{fig05} Time-reversibility test on EEG data. Insets show ten seconds in the same scales (a) before and (b) during the epileptic episode. Same stipulations as in the caption of Fig.~\ref{fig04}
}
\end{figure}

As many others time series in biology and medicine, electroencephalographic (EEG) signals display strong nonlinearities during different cognitive or pathological states~\cite{Stam1998}. Time-reversibility can be a useful property of interictal EEG signals, as it can serve as a marker of the epileptogenic zone~\cite{Heyden1996, Pijn1997, Schindler2016}. Here, we applied our TiROP test to scalp EEG recordings from a pediatric subject with intractable epileptic seizures~\cite{EEGdatabase2000, Goldberger2000, Shoeb2009}.  Figs. \ref{fig05}(a)-(b) show the time series corresponding to the interictal and ictal (seizure) periods, respectively.  Our results confirm previous findings suggesting that interictal EEG dynamics can be associate to a reversible linear process, whereas time irreversibility characterizes epileptic seizures~\cite{Heyden1996, Pijn1997, Schindler2016}. 

\section{Conclusions}
In this work we have addressed the problem of detecting, from scalar observations, the time scales involved in temporal irreversibility. Based on the ordinal patterns analysis, the TiROP method compares the information content of the symbolic representation of $X_t$  and the counterpart of its time-reversed version $X'_t$. 
In contrast with other approaches based on symbolic analysis, the approach proposed here has the key practical advantage that it is fully data-driven and it does not require any \textit{a priori} thresholds, or any knowledge about the data sequence for its symbolic representation, which is very useful in real-world data analysis. 

Results confirm that TiROP provides an interesting and promising approach to the analysis of complex time series. The applicability and advantages of our method was demonstrated by many examples from synthetic and real, linear and nonlinear models. The method outperforms a classical moment-based test, which often fails to detect time-irreversibility along different time-lags. Our results confirm temporal irreversibility in economical time series, and suggest this property as a common signature in epidemiological data. This would imply that  additional nonlinear analysis techniques should be applied for a more complete characterization of such time series. The results indicates that time irreversibility can also be observed at scalp EEG recordings of epileptic seizures in humans. 

To conclude, this study shows that the detection of temporal irreversibility in time series can be successfully addressed using ordinal symbolic representation. The main advantage of our proposal relies on its simplicity, reliability and computational efficiency thanks to the ordinal patterns transformation and analysis. The detection of temporal irreversibility in other data (e.g. cardiac or climate time series) might provide meaningful insights into the underlying process generating the observed time series. This framework could also add new functionality to current non-linear analysis techniques, but also it could open the way to define physiological biomarkers.

\begin{acknowledgments}
JHM and MC are grateful to members of Gnonga-Tech for useful and valuable suggestions. JLHD is supported by the S\~ao Paulo Research Foundation (FAPESP) under grants 2016/01343-7 and 2017/00344-2.
\end{acknowledgments}

\end{document}